\documentclass{amsart}
\usepackage{cite}
\usepackage{a4wide}

\usepackage[british,UKenglish,USenglish,english,american]{babel}

\usepackage{amssymb,amsmath,amsthm,enumerate,amsfonts}
\usepackage{newlfont}
\usepackage{dsfont}
\usepackage{amsbsy}
\usepackage{graphicx}

\def\CC{\mathrm \hbox{C\hspace{-1.em}\raisebox{.47ex}{
         \textrm{\mbox{$\scriptscriptstyle |$}}}\hspace{+0.8 em} }}
\def\CC{\hskip.2em\raisebox{.22em}{\makebox[0pt][c]{$\scriptscriptstyle |$}}
\hskip-0.25em\mathrm{C}}       

\def\hcboxcm#1#2{\hbox to #1{\hfill #2 \hfill}}

\def\null{\hbox{}}

\def\tn1{\widetilde n_1}
\def\tn2{\widetilde n_2}
\def\tn{\widetilde n }

\let\ds\displaystyle

\def\be{\begin{equation}}
\def\ee{\end{equation}}
\def\bea{\begin{eqnarray}}
\def\eea{\end{eqnarray}}

\def\bean{\begin{eqnarray*}}
\def\eean{\end{eqnarray*}}

\def\={\, = \, }

 \def\OO{\rm \hbox{O\kern-.34em\raise.47ex
         \hbox{$\scriptscriptstyle |$}\kern-.46em\raise.47ex
         \hbox{$\scriptscriptstyle |$}\kern+0.5 em }}
\def\RR{{\mathbb R} }

\def\F{{{\cal F}}}

\def\CC{\mathbb{C}}

\def\OO{\mathbb{O}}

\def\RR{\mathbb{R}}

\def\Box{\leavevmode\vbox{\hrule
     \hbox{\vrule\kern4pt\vbox{\kern4pt}%
           \vrule}\hrule}}
\def\blackbox{\leavevmode\vrule height 5pt width 4pt depth 0pt\relax}
\catcode`@=11

\def\eqalign#1{\null\,\vcenter{\openup1\jot \m@th
   \ialign{\strut \hfil$\displaystyle{##}$ & $\displaystyle{{}##}$\hfil
      \crcr#1\crcr}}\,}
\def\eqalignrll#1{\null\,\vcenter{\openup1\jot \m@th
   \ialign{\strut \hfil$\displaystyle{##}$ & $\displaystyle{{}##}$\hfil
    & $\displaystyle{{}##}$\hfil
      \crcr#1\crcr}}\,}
\def\eqalignrcl#1{\null\,\vcenter{\openup1\jot \m@th
   \ialign{\strut \hfil$\displaystyle{##}$ &\hfil $\displaystyle{{}##}$\hfil
    & $\displaystyle{{}##}$\hfil
      \crcr#1\crcr}}\,}
\def\eqalignno#1{\displ@y \tabskip\@centering
  \halign to\displaywidth{\hfil$\@lign\displaystyle{##}$\tabskip\z@skip
    &$\@lign\displaystyle{{}##}$\hfil\tabskip\@centering
    &\llap{$\@lign##$}\tabskip\z@skip\crcr
    #1\crcr}}
\newcounter{appendix}
\newcounter{sectionz}
\def\appendix{\advance\c@appendix by 1
\def\thesectionz {\Alph{appendix}}
\def\thesection{\Alph{appendix}} 
   \ifnum\c@appendix=1 \setcounter{section}{-1} \fi
   \@startsection {section}{1}{\z@}{-3.5ex plus -1ex minus 
  -.2ex}{2.3ex plus .2ex}{\large\bf}}

\catcode`@=12
\newtheorem{lemme}{Lemma}[section]  

\newtheorem{theorem}[lemme]{Theorem}

\newtheorem{corollary}[lemme]{Corollary}

\newtheorem{definition}[lemme]{Definition}

\newtheorem{proposition}[lemme]{Proposition}

\newtheorem{remark}[lemme]{Remark} 

\def\deblem{\begin{lemme}\it }
\def\finlem{\end{lemme}}
\def\debthm{\begin{theorem}\it }
\def\finthm{\end{theorem}}
\def\debprop{\begin{proposition} \it}
\def\finprop{\end{proposition}}
\def\debcor{\begin{corollary}\it }
\def\fincor{\end{corollary}}
\def\debdef{\begin{definition}\it}
\def\findef{\end{definition}}
\def\debrem{\begin{remark}\em}
\def\finrem{\null\hfill\blackbox\end{remark}}
\begin{document}
\title[Solvable models of quantum beating]{Solvable models of quantum beating}

\author[R.~Carlone,\,R.~Figari,\,C.~Negulescu,\,L. ~Tentarelli ]{$^{1}$R.~Carlone,$^{2}$R.~Figari,$^{3}$C.~Negulescu,$^{4}$L.~Tentarelli }
\address{$^{1}$Universit\`{a} ``Federico II'' di Napoli, \\Dipartimento di Matematica e Applicazioni ``R. Caccioppoli'',\\ MSA, via Cinthia, I-80126, Napoli, Italy\\
$^{2}$Universit\`{a} ``Federico II'' di Napoli, \\ Dipartimento di Fisica e INFN Sezione di Napoli,\\MSA, via Cinthia, I-80126, Napoli, Italy\\
$^{3}$Universit\'e de Toulouse \& CNRS, UPS\\ Institut de Math\'ematiques de Toulouse UMR 5219\\F-31062 Toulouse, France\\
$^4$ Sapienza Universit\`a di Roma,\\ Dipartimento di Matematica ,\\ Piazzale Aldo Moro, 5, 00185, Roma, Italy}
\email{raffaele.carlone@unina.it, rodolfo.figari@na.infn.it, claudia.negulescu@math.univ-toulouse.fr, tentarelli@mat.uniroma1.it}

\begin{abstract}
We review some results about the suppression of quantum beating in a one dimensional non-linear double well  potential. We implement a single particle double well potential model making use  of nonlinear point interactions.  We show that there is  complete suppression of the typical beating phenomenon characterizing the linear quantum case.
\end{abstract}

\noindent

\maketitle



\section[Introduction]{Introduction}

In the last decades the quantum beating phenomenon has become a subject of great interest in different areas of  quantum physics, ranging from quantum electrodynamics to particle physics, from solid state physics to molecular structure and dynamics.

\begin{figure}[htbp]
    \centering
\includegraphics[width=6cm, angle =-90]{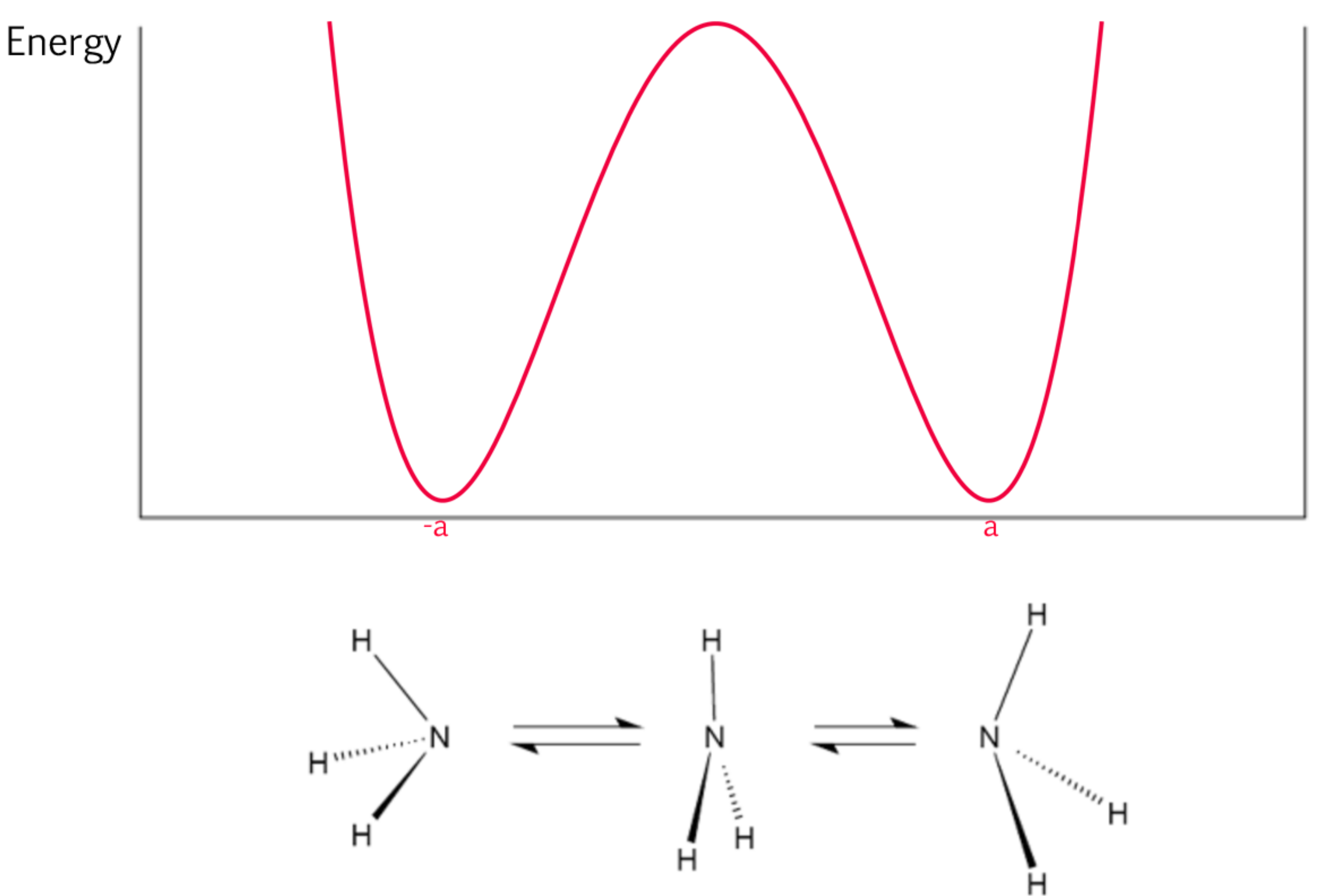}
 \caption{Schematical representation of Ammonia molecule.}
    \label{inv}
\end{figure}

Quantum beating was first experimentally observed in 1935 as a periodic inversion of the nitrogen atom with respect to the hydrogen atoms plane in the ammonia molecule. The phenomenon was then theoretically investigated examining the one dimensional dynamics of a quantum particle in a double well potential, the simplest example of a bistable potential. In figure (\ref{inv}) the two minima correspond to the average positions of the nitrogen atom in the two symmetric states.

The ammonia molecule is pyramidally shaped. Three hydrogen atoms form the base and the nitrogen atom is located in one of the two distinguishable  states (enantiomers) on one side or the other with respect to the base (chirality) 
. The experiments on liquid ammonia showed that a microwave radiation could  induce a periodic transition from one state to the other (quantum beating). It was also observed that the periodic nitrogen inversion was absent whenever the molecule was part of a large organic structure or the pressure was too high.

Many authors used an effective non-linear potential, superimposed to the double well, to model the interaction of the single molecule with the outside structure (see \cite{Davies:1979ep,Davies:1995gb,JonaLasinio:2001gr,Herbauts:2007ix}).

The beating phenomenon for a particle in a double well potential is expected to be visible when the ground state and the first excited state have very close energies, forming an almost single, degenerate, energy level. A superposition of these two states will evolve concentrating  periodically inside one well or the other, with a frequency proportional to the energy difference (see section \ref{SEC22} below).

When a nonlinear effective potential is assumed to model the interaction with the environment the dynamics to be investigated is the nonlinear Schr\"odinger equation 

\[
\imath\hbar\frac{\partial\psi}{\partial\,t}=-\frac{\hbar^{2}}{2\,m}\frac{\partial^{2}\psi}{\partial\,x^{2}}+V(x)+\varepsilon|\psi|^{\sigma}\psi
\]
where $V(x)$ is a double well potential.

In order to comprehend the beating suppression induced by the environment one needs to prove  
that the non linear interaction destroys the periodic dynamics  for all initial conditions, which in turn implies that the particle will be eventually confined in one of the wells (in which well the particle will finally collapse will depend on the specific chosen nonlinearity and/or on the initial conditions).

In a completely symmetric nonlinear double well Sacchetti proved (see \cite{S04}) that for $\sigma = 2$ the result holds true, in the semiclassical approximation (i.e., ``$\hbar \rightarrow 0$''). Related results were obtained in \cite{Grecchi:1996de,Grecchi:2002bh,Sacchetti:2004fp,Sacchetti:2006io} in any dimension assuming a symmetry breaking nonlinear perturbation of the double well potential.\\

In \cite{CFN17} the beating phenomenon in the case of a linear and nonlinear ``point well interactions'' (for more, refer to \cite{SAlbeverio:2011vta}) is analyzed. The main advantage in using point interactions is that explicit solutions for the linear dynamics are available. Moreover, the analysis of the Schr\"odinger equation in the case of a nonlinear point interaction hamiltonian can be reduced to the search of the solutions to a system of nonlinear Volterra integral equations for a complex function depending only on time. At least at the level of numerical computation this reduction turns out to be a remarkable simplification.

In this review we recall first definition and properties of linear and nonlinear point interaction hamiltonians.
In the successive section we consider a one dimensional hamiltonian with two attractive point potentials. We examine its spectral properties and characterize the dependence of the energy difference between the ground state and the first exited state on the kinematical and dynamical parameters of the interaction. In this way, we will then be able to write down explicitly the beating solution for any range of parameters and successively to investigate the semiclassical limit.

We then investigate the Cauchy problem for the Schr\"odinger equation with two nonlinear point well potentials. As already mentioned, the description of the dynamics will be reduced to the analysis of a system of two Volterra integral equations. Based mainly on numerical analysis results we will discuss the beating suppression. 

\section{The mathematical model - Concentrated nonlinearities} 

First, we briefly recall the definition of point interaction hamiltonians in $L^{2}(\mathbb{R})$ (see \cite{SAlbeverio:2011vta} for further details). For two point scatterers placed in $Y=\{y_{1},y_{2}\}$ of strength  $\underline{\gamma} = \{\gamma_{1}, \gamma_{2}\} ,\,\,\, y_i,\,\gamma_{i} \in \mathbb{R}$, the formal hamiltonian reads 
\begin{equation} \label{formal}
H_{\underline{\gamma}, Y}\, 
\psi :=`` - \frac{d^{2}}{dx^{2}} \, \psi + \gamma_1 \delta_{y_1} \psi + \gamma_2 \delta_{y_2} \psi\,``,
\end{equation}
where the (reduced) Planck constant $\hbar$ has been taken equal to one  and the particle mass $m$ equal to $1/2$.
We will also assume that the two points are placed symmetrically with respect to the origin and that $|y_{i}|=a$.  \\ 

The following result holds true (see \cite{SAlbeverio:2011vta}):
\begin{equation} \label{dom}
D(H_{\underline{\gamma}, Y}) := \Big\{ \psi \in L^{2}(\mathbb{R}) \;| \; \psi =
\phi^{\lambda} \,-\,   \sum_{i,j=1}^{2} \left( \Gamma^{\lambda}_{\underline{\gamma}} \right)^{-1}_{ij} \phi^{\lambda}(y_{j}) G^{\lambda}
(\cdot - y_{i}), \; \phi^{\lambda} \in H^{2}(\mathbb{R}) \Big\} \,,
\end{equation}
\be\label{oper}
\left(H_{\underline{\gamma}, Y} + \lambda\right) \psi = \left(-  \frac{d^{2}}{dx^{2}} + \lambda\right) \phi^{\lambda},
\ee
are domain and action of a selfadjoint operator in $L^{2}(\mathbb{R})$ which acts as  the free laplacian on functions supported outside the two points $y_{i}=\pm a$. In (\ref{dom}) $G^{\lambda}(\cdot)$ is the free laplacian Green function
\[
G^{\lambda}(x):=\frac{e^{-\sqrt{\lambda}|x|}}{2\sqrt{\lambda}},
\]
and the matrix $\Gamma^{\lambda}_{\underline{\gamma}}$ is defined as
\[
 \left( \Gamma^{\lambda}_{\underline{\gamma}} \right)_{ij}\ :=  \frac{1}{\gamma_{i}}  \, \delta_{ij}  + G^{\lambda} (y_{i} - y_{j})\,,
\]
where the positive real number $\lambda$ is chosen large enough to make the matrix $\Gamma^{\lambda}_{\underline{\gamma}}$ invertible.

It is immediate to check that the derivative of $G^{\lambda}(x)$ has a jump in the origin, equal to $-1$. This in turn implies that every function $\psi\,$ in the domain satisfies the boundary conditions  
\be \label{bc}
\frac{d \psi}{dx}\left(y_{j}^{+}\right) -\frac{d \psi}{dx}\left(y_{j}^{-}\right)  = \gamma_{j} \,\psi (y_{j})\,, \quad j=1,2\,.
\ee
The dynamics generated by $H_{\underline{\gamma}, Y}$ is then characterized as the free dynamics outside the two scatterers, satisfying at any time the boundary conditions (\ref{bc}).\\
 
Our aim is to investigate the behaviour of the solutions of the nonlinear evolution problem
\begin{equation}\label{SCH}
\left\{
\begin{array}{l}
\ds {\imath}\, \frac{\partial \psi}{\partial t}  = H_{\underline{\gamma} (t), Y}\, \psi\,, \quad \forall (t,x) \in \RR^+ \times \RR\,,\\[3mm]
\ds \psi(0,x)=\psi_0(x) \in D(H_{\underline{\gamma}(0), Y})\quad \forall x \in \RR\,, \\[3mm]
\ds \gamma_j(t):= \gamma |\psi(t,y_j)|^{2 \sigma}, \,\,\,\gamma < 0, \,\,\, \sigma \geq 0.
\end{array}
\right.
\end{equation}
where the time dependence of $\underline{\gamma}$ is non-linearly determined by the values in $ \pm a$ of the solution itself.

There is an alternative way to represent the solutions of the Cauchy problem (\ref{SCH}). Let us consider the following ansatz, suggested by the Duhamel's formula applied to the evolution equation (\ref{SCH}) using the formal definition (\ref{formal}) for the Hamiltonian,
\begin{equation}\label{duhamel_0}
\psi(t,x)=(\mathcal{U}(t)\psi_{0})(x)-{\imath}\,\gamma\sum_{j=1}^{2}\int_{0}^{t} U(t-s;x-y_{j})|\psi(s,y_{j})|^{2\sigma}\psi(s,y_{j})\, ds\,.
\end{equation}
where $U(\tau, y)$ is the integral kernel of the unitary group $\displaystyle e^{\imath t \Delta}$, i.e.
$$
U(\tau,y):=\frac{e^{\imath\frac{|y|^2}{4\tau}}}{\sqrt{4 {\imath}\, \pi\, \tau}}\,, \qquad (\mathcal{U}(t)\xi)(x)= \int_{-\infty}^\infty U(t;x-y)\, \xi(y)\, dy,\quad \forall \xi \in L^{2}(\RR).
$$
From ansatz (\ref{duhamel_0}) one obtains for $i=1,2$
\[
\psi(t,y_{i})=(\mathcal{U}(t)\psi_{0})(y_{i})-{\imath}\,\gamma\sum_{j=1}^{2}\int_{0}^{t} U(t-s;y_{i}-y_{j})|\psi(s,y_{j})|^{2\sigma}\psi(s,y_{j})\, ds.
\]
Explicitly 
\be \label{VOLT}
\left\{
\begin{array}{l}
\ds \psi(t,-a)+\frac{\gamma}{2} \sqrt{\frac{\imath}{\pi}} \, \int_0^t  \frac{ \psi(s,-a)\, |\psi(s,-a)|^{2 \sigma}}{\sqrt{t-s}}\, ds + \\[.7cm] 
\ds \hspace{3cm} + \frac{\gamma}{2} \sqrt{\frac{\imath}{\pi}} \, \int_0^t  \frac{ \psi(s,a)\, |\psi(s,a)|^{2 \sigma}}{\sqrt{t-s}} \, e^{\imath\frac{ a^2}{(t-s)}}\, ds = ({\mathcal U}(t)\, \psi_0)(-a)\,,\\[1.2cm]
\ds \psi(t,a)+\frac{\gamma}{2} \sqrt{\frac{\imath}{\pi}} \, \int_0^t  \frac{ \psi(s,a)\, |\psi(s,a)|^{2 \sigma}}{\sqrt{t-s}}\, ds + \\[7mm]
\ds \hspace{3cm} \frac{\gamma}{2} \sqrt{\frac{\imath}{\pi}} \, \int_0^t  \frac{ \psi(s,-a)\, |\psi(s,-a)|^{2 \sigma}}{\sqrt{t-s}} \, e^{\imath\frac{ a^2}{(t-s)}}\, ds = ({\mathcal U}(t)\, \psi_0)(a)\,.
\end{array}
\right.
\ee

\medskip
\noindent The problem was extensively discussed in \cite{Adami:2001bt}, where it was proved that, if $\psi (t, \pm a)$ are solutions of (\ref{VOLT}), then  the function  (\ref{duhamel_0}) is the unique solution of \eqref{SCH} (see \cite{CCF17,CFT17,CCT17} and \cite{ADFT03,ADFT04} for d=2 and d=3). 

\begin{remark}
 It is worth pointing out that the solution of \eqref{SCH} mentioned before is guaranteed to be \emph{global-in-time} only if $\sigma<1$. On the other hand, whenever $\sigma\geq1$ there exist initial data for which blow-up phoenomena may arise.
\end{remark}

\begin{remark}
 Throughout, we use the notation $q_{1}(t)  \equiv \psi(t,-a) , \,\,q_{2}(t)  \equiv \psi(t,a)$ and refer to (\ref{VOLT}) as the ``charge equations''. 
\end{remark}

In the following subsection we examine the linear case analysing the necessary conditions to have quantum beating states.

\subsection{Linear point interactions} 

Let us consider the linear case, corresponding to $\sigma=0$ and $\gamma_{j} < 0$, for $j=1,2$, independent of $t$ in (\ref{SCH}). From the definition (\ref{oper}) the resolvent of the operator $H_{\underline{\gamma}, Y}$ has integral kernel 
\be\label{respi}
(H_{\underline{\gamma}, Y} +\lambda)^{-1} (x,x')= G^{\lambda}(x - x') - \sum_{i,j =1}^2  \left( \Gamma^{\lambda}_{\underline{\gamma}} \right)^{-1}_{ij} G^{\lambda}(x - y_i) G^{\lambda} (x' - y_j).
\ee
As it is clear from (\ref{respi}),  $H_{\underline{\gamma}, Y}$  is a finite rank perturbation of the free laplacian resolvent operator. This in turn implies that the essential spectrum of $H_{\underline{\gamma}, Y}$ is $[0,\infty)$ and that $-\lambda$ is a negative eigenvalue if and only if the matrix $\Gamma^{\lambda}_{\underline{\gamma}} $ is not invertible 

\[
\det\Gamma_{(\gamma_{1},\gamma_{2})}^\lambda= \det\, \left(\begin{array}{cc}\frac{1}{\gamma_{1}}+\frac{1}{2\,\sqrt{\lambda}}& G^{\lambda}(2 a) \\ G^{\lambda}(2 a)  & \frac{1}{\gamma_{2}} + \frac{1}{2\,\sqrt{\lambda}}\end{array}\right)=0\,,
\]
or
\begin{equation}\label{eigen}
\left(\frac{1}{\gamma_{1}}+\frac{1}{2\,\sqrt{\lambda}}\right)\left(\frac{1}{\gamma_{2}}+\frac{1}{2\,\sqrt{\lambda}}\right)-\left(\frac{1}{2\,\sqrt{\lambda}}\right)^{2}e^{- 4 \sqrt\lambda \, a}=0\,.
\end{equation}

All the relevant results about the point spectrum of  $H_{\underline{\gamma}, Y} $ are collected in the following lemma.  

\begin{lemme}[\cite{CFN17}] \label{lemmino}
Let  $\gamma_1\leq \gamma_2 $ and let us define the ratio $\displaystyle \alpha := \frac{\gamma_{2}}{\gamma_{1}} $. Then one has:
\begin{description}
\item[a)] There are two real solutions $\lambda_{0} > \lambda_{1}>0$ to equation (\ref{eigen}) if and only if $\gamma_{i} < 0$ for $i=1,2$ and
\begin{equation}\label{two}
\frac{1}{|\gamma_{1}|}+\frac{1}{|\gamma_{2}|}< 2a\,.
\end{equation}
\item[b)]
For $\gamma_{i} = \gamma < 0$, $i=1,2$, satisfying \eqref{two} ($1/\gamma < a$) , one has
\[\Delta \lambda := \lambda_{0} - \lambda_{1}   \simeq \gamma^{2} e^{-|\gamma| \alpha}\]
In particular $\Delta \lambda \rightarrow 0$ exponentially as $|\gamma| \alpha \rightarrow \infty$\,.

\item[c)]
For $\gamma_{i} < 0$, $i=1,2$, satisfying \eqref{two} and $\alpha <1$, one has
\[\Delta \lambda := \lambda_{0} - \lambda_{1} \geq \gamma_{1}^{2}  (1 -\alpha^{2})\,. \]
In particular $\Delta \lambda \rightarrow \infty$ as $|\gamma_{1}| \rightarrow \infty$\,.

\vspace{.1cm}
\item[d)]  For $\gamma_{i} < 0$, $i=1,2$, satisfying \eqref{two} and $\alpha \leq1$, one has
\[\lim_ {|\gamma_{1}| \rightarrow \infty}    2 \sqrt{\lambda_{0}} / \gamma_{1} = -1\,, \quad \lim_ {|\gamma_{1}| \rightarrow \infty} 2 \sqrt{\lambda_{1}} /\gamma_{2} = -1\,.\]

\item[e)]  For $\gamma_{i} < 0$, $i=1,2$, satisfying \eqref{two}, the eigenfunctions associated with the two negative eigenvalues are (see  \cite{SAlbeverio:2011vta}) 
\begin{equation}\label{eigenfunction0}
\phi_0(x)=c_{0}G^{\lambda_{0}}(x-y_{1})+c_{1\,}G^{\lambda_{0}}(x-y_{2})\,,
\end{equation}
\begin{equation}\label{eigenfunction1}
\phi_1(x)=c_{2} G^{\lambda_{1}}(x-y_{1})+c_{3}G^{\lambda_{1}}(x-y_{2})\,, 
\end{equation}
where the coefficients $c_{0},c_{1}$ and $c_{2},c_{3}$ are solutions of
\begin{equation}\left(\begin{array}{cc}\frac{1}{\gamma_{1}}+\frac{1}{2\,\sqrt{\lambda_0}}& \frac{1}{2\sqrt{\lambda_0}}e^{-2\,\sqrt{\lambda_0}\,a} \\ \frac{1}{2\sqrt{\lambda_0}}e^{-2\,\sqrt{\lambda_0}\,a}  & \frac{1}{\gamma_{2}} + \frac{1}{2\,\sqrt{\lambda_0}}\end{array}\right)\left(\begin{array}{c}c_{0}\\c_{1}\end{array}\right)=\left(\begin{array}{c}0\\0\end{array}\right)\,, \nonumber
\end{equation}
and 
\begin{equation}\left(\begin{array}{cc}\frac{1}{\gamma_{1}}+\frac{1}{2\,\sqrt{\lambda_1}}& \frac{1}{2\sqrt{\lambda_1}}e^{-2\,\sqrt{\lambda_1}\,a} \\ \frac{1}{2\sqrt{\lambda_1}}e^{-2\,\sqrt{\lambda_1}\,a}  & \frac{1}{\gamma_{2}} + \frac{1}{2\,\sqrt{\lambda_1}}\end{array}\right)\left(\begin{array}{c}c_{2}\\c_{3}\end{array}\right)=\left(\begin{array}{c}0\\0\end{array}\right)\,. \nonumber
\end{equation}

\end{description}
\end{lemme}

\vspace{.5cm}
Solving explicitly the last equations at point {\bf{e)}} of Lemma \ref {lemmino} we obtain
\[
\left|\frac{c_{1}}{c_{0}}\right| =\sqrt{\frac{(2\sqrt{\lambda_{0}}/ \gamma_{1}) +1}{
(2\sqrt{\lambda_{0}}/ \gamma_{2}) +1}}\,, \nonumber
\]
\[
\left| \frac{c_{2}}{c_{3}} \right |=\, \sqrt{\frac{(2\sqrt{\lambda_{1}}/ \gamma_{2}) +1}{
(2\sqrt{\lambda_{1}}/ \gamma_{1}) +1
}}\,. \nonumber
\]
The normalization condition finally gives
\begin{equation}\label{czero}
c_{0}=\frac{2 |\gamma _1| \lambda _0^{3/4}}
{\sqrt{\gamma _1\gamma _2\frac{\left(\gamma _1+2 \sqrt{\lambda _0}\right)}{\left(\gamma _2+2 \sqrt{\lambda _0}\right)} +\gamma _1 \left( \gamma _1+4 \sqrt{\lambda _0} +2 \sqrt{\lambda _0}\, a\left(\gamma _1 +2 \sqrt{\lambda _0}\right)\right)}}\,,
\end{equation}
\begin{equation}\label{ctre}
c_{3}=\frac{2 |\gamma _2| \lambda _1^{3/4}}
{\sqrt{\gamma _1\gamma _2\frac{\left(\gamma _2+2 \sqrt{\lambda _1}\right)}{\left(\gamma _1+2 \sqrt{\lambda _1}\right)} -\gamma _2 \left( \gamma _2+4 \sqrt{\lambda _1} +2 \sqrt{\lambda _1}\, a\left(\gamma _2 +2 \sqrt{\lambda _1}\right)\right)}}\,. 
\end{equation}
Few remarks are worth doing:
\begin{remark}
In our units the condition characterizing the semi-classical limit is $\delta := |\gamma| \alpha >> 1$. In standard units the condition would be $ \bar{\delta} := \displaystyle \frac{2m |\gamma| a}{\hbar^2} >>1$ and the energy difference 
\begin{equation}
\triangle E \simeq \frac{2\,m\,\gamma^{2}}{\hbar^{2}}e^{-\bar{\delta}}\,. \nonumber
\end{equation}
\end{remark}

\begin{remark}
Notice the extreme instability of the energy difference with respect to the ratio $\alpha$ when it is closed to the value one. While in the symmetric case ($\alpha = 1$) the energy difference is decreasing exponentially in the semiclassical limit the same quantity is going to infinity in the same limit for $\alpha < 1$. This fact will appear to be the main reason in the quantum beating suppression in the asymmetric and in the non linear case.
\end{remark}
\begin{remark} \label{coefratio}
In the semi-classical limit, the coefficient ratios (\ref{czero}) and (\ref{ctre})  tend to $1$ in the symmetric case ($\alpha =1$) whereas they tend to $0$ for any $\alpha <1$. In turn this means that, in the same limit,  the eigenfunctions (\ref{eigenfunction0}) and (\ref{eigenfunction1}) tend to be equally distributed on the two wells if $\alpha =1$ whereas they are strongly confined in one of the well for any $\alpha < 1$.

\end{remark}
\subsection{The beating phenomenon } \label{SEC22}

Now, let us consider the linear case when the condition $\frac{1}{|\gamma_{1}|}+\frac{1}{|\gamma_{2}|}< 2a$ for the existence of two eigenvalues is fulfilled. Following a standard notation, we will use in this subsection subscripts $"f,e"$ instead of $0,1$ to mean ``\underline{fundamental}'' and ``first \underline{excited} state'' (respectively). The corresponding eigenfunctions are 

\[
\phi_{f}(x)=c_{0 }G^{\lambda_f}(x+a)+ c_{1}G^{\lambda_f}(x-a),
\]
\[
\phi_{e}(x)=c_{2} G^{\lambda_{e}}(x+a)- c_{3} G^{\lambda_{e}}(x-a)\,.
\]
The superposition of the two eigenfunctions
\[
\psi^{L}_{beat,0}(x):=\frac{1}{\sqrt{2}}\left(\phi_{f}(x)+\phi_{e}(x)\right)\,
\]
will evolve in time as follows:

\begin{equation}\label{lbeat0}
\psi_{beat}^{L}(t,x) =\frac{1}{\sqrt{2}}\left(e^{{\imath}\lambda_f t}\phi_{f}(x)+e^{{\imath}\lambda_e t}\phi_{e}(x)\right)\,,
\end{equation}
with a probability density  given by
\[
\mathcal{P}(t,x)=\frac{1}{2}\left[|\phi_{f}(x)|^{2}+|\phi_{e}(x)|^{2}+2\, \phi_{f}(x)\phi_{e}(x)\cos\left( (\lambda_f-\lambda_e)t\right) \right]\,. 
\]

\begin{figure}[htbp]
    \centering
\includegraphics[width=8cm]{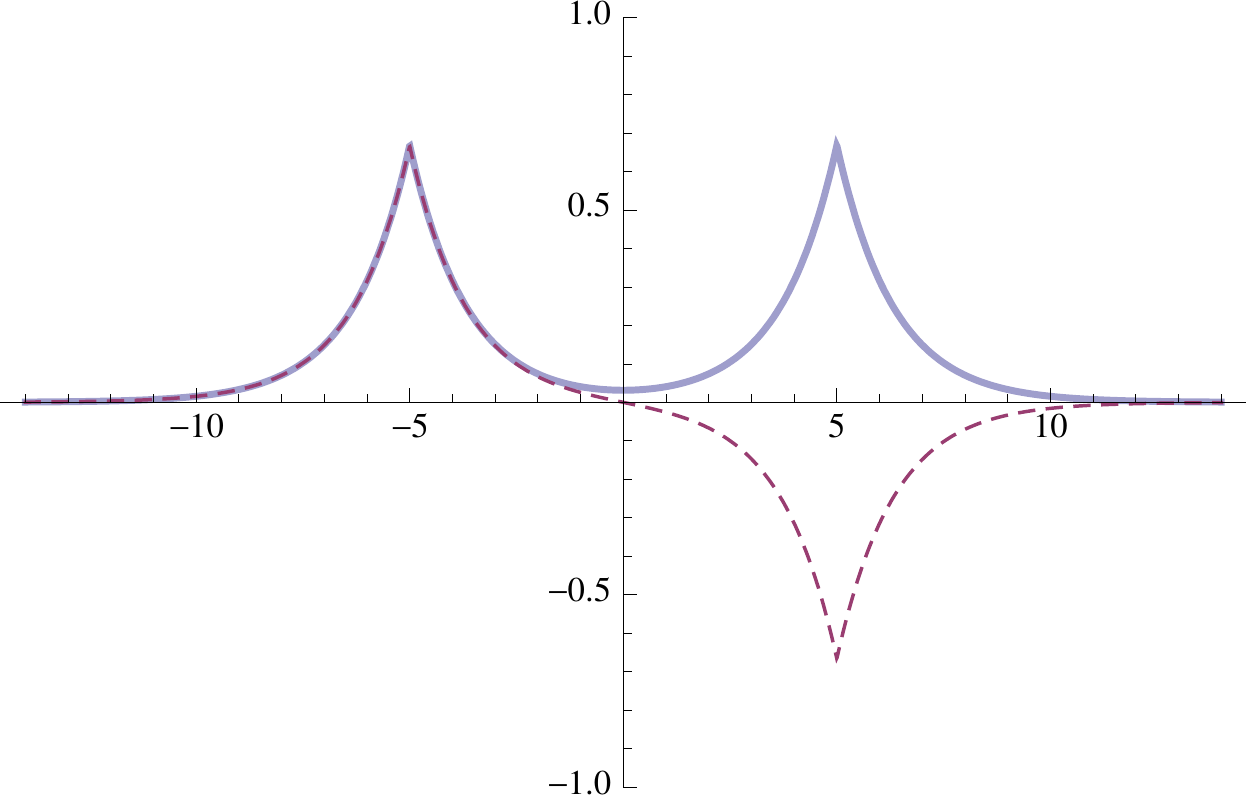}
 \caption{Plot of the functions $\phi_{f}(x)$ with a  thicked blue line and $\phi_{e}(x)$ with a dashed line.}
    \label{twoen}
\end{figure}

Let us consider first the symmetric case. The two eigenfunctions $\phi_{f}$ and  $\phi_{e}$ are respectively symmetric and antisymmetric with respect to the origin and have similar absolute value everywhere (see Figure (\ref{twoen}))

\noindent 
It is clear that $\psi_{beat}^{L}$ is initially supported around the point $-a$ and is an oscillating function with period $\displaystyle T_B=\frac{2\pi}{|\lambda_f-\lambda_e|} $  concentrated periodically on the left and on the right well, justifying the definition of (\ref{lbeat0}) as a beating state.

\begin{figure}[htbp]
    \centering
\includegraphics[width=10cm]{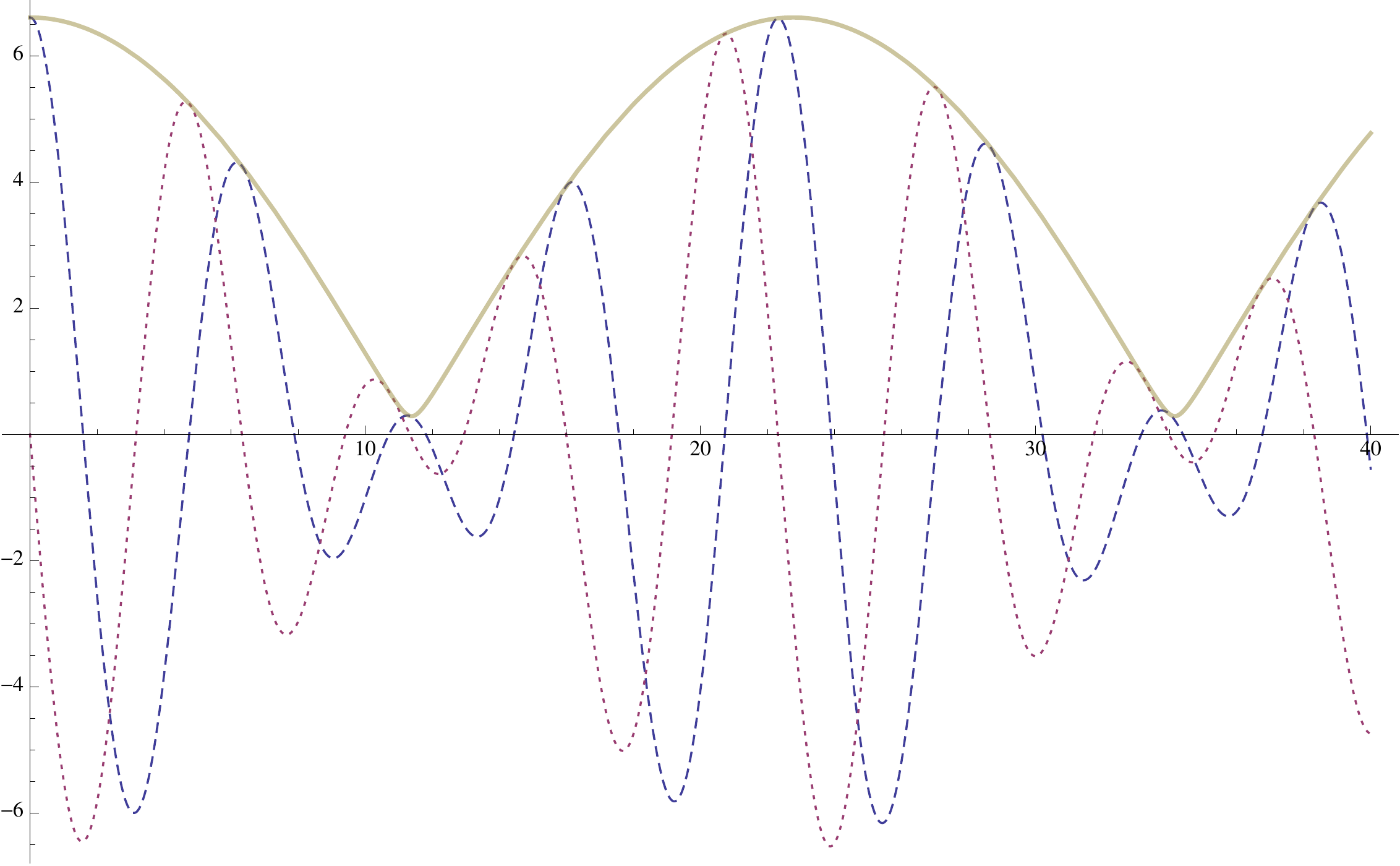}
 \caption{Plot of the time-evolution of the functions $\textrm{Re}\,q^{L}_{1}(t)$ as a dashed line, $\textrm{Im}\,q^{L}_{1}(t)$ as a dotted line and $|q^{L}_{1}|(t)$ as a thick line.}
    \label{q}
\end{figure}

The values assumed by the function $\psi_{beat}^{L}(t,x)$ in the centers of the two wells evolve as follows 
\[
\begin{array}{lll}
q_{1}^{L}(t) &\equiv&\ds  \psi_{beat}^{L}(t,-a)=\frac{1}{\sqrt{2}}\left(e^{{\imath}\lambda_f t}\phi_{f}(-a)+e^{{\imath}\lambda_e t}\phi_{e}(-a)\right)\\[3mm]
q_{2}^{L}(t) &\equiv&\ds \psi_{beat}^{L}(t,a)=\frac{1}{\sqrt{2}}\left(e^{{\imath}\lambda_f t}\phi_{f}(a)+e^{{\imath}\lambda_e t}\phi_{e}(a)\right)\,.
\end{array}
\]
and are plotted in figure \ref{q}.
 
\vspace{.5cm}
The situation is remarkably different when $\alpha <1$. In this case, as we pointed out in Remark \ref{coefratio}, the two eigenstates are strongly confined in different wells for $\gamma_{1}$ large. In particular their product is going to be zero almost everywhere. Any initial superposition of the two eigenstate
\[
\psi_{asy,0}(x):= \alpha\, \phi_0(x)+\beta\, \phi_1(x)\,, \quad \alpha,\beta \in \CC\,,\,\,\,\,\,|\alpha|^2 + |\beta|^2 = 1
\]
will evolve at time $t$ into the state
 \[
 \psi_{asy}(t,x):= \alpha\, e^{\imath \, \lambda_0\, t}\, \phi_0(x)+\beta\,e^{\imath \, \lambda_1\, t}\, \phi_1(x)\,.
 \]
but the time dependent term in the square modulus will be negligible and no beating phenomenon will take place. 

One expects that the asymmetry due to the non-linearity will produce a similar suppression on time scales depending on the initial condition and on the strength of the nonlinearity. 

\subsection{Nonlinear point interactions} 

A detailed analytical study of the non-linear case $\sigma >0$ can be found in  \cite{Adami:1999tk,Adami:2001bt} where the authors prove general results about existence of solutions, either local or global in time, and prove existence of blow-up solutions for $\sigma \geq 1$. 
 
In the following we  analyze  results about the evolution of a beating state obtained via numerical computation. A complete analytical analysis of equation (\ref{VOLT}) is still lacking. The problem is to quantify the amount of asymmetry necessary to suppress quantum beating induced  by the nonlinearity and the time elapsed before that level is reached.

Let us consider an initial state which would evolve in a quantum beating state in the linear case
\[ 
\psi_0 (x):=\alpha \,\phi_{f}(x)+ \beta \, \phi_{e}(x)\,, \quad \alpha,\beta \in \CC\,\,\,\,\, |\alpha|^2 + |\beta|^2 = 1
\]
In the following we investigate the Cauchy problem (\ref{SCH}) with initial conditions 
\[
 \psi(0,x)=\psi_0(x)
\]
using its integral form (\ref{VOLT}). From \cite[Theorem 6]{Adami:2001bt}, we know that, under the assumptions $\sigma < 1$ and $\psi_0 \in H^{1}(\mathbb{R})$, the Cauchy problem has a unique solution which is global in time. Moreover in \cite[Theorem 23]{Adami:2001bt}, it is proved that if $\gamma<0$ and $\sigma\geqslant 1$ then there exist initial data such that the solutions of the Cauchy problem  blow-up in finite time.

Let us assume now that $\gamma<0$ and $\sigma< 1$. We list the solutions to (\ref{VOLT}) obtained by numerical computation in \cite{CFN17}. In particular, we will compare the solution in the linear case with solutions to (\ref{VOLT}) with increasing powers of the non linearity. Our results show how the asymmetry generated by the nonlinear interactions produce  the  complete suppression of the beating phenomenon. 
\vspace{0.5cm}

For the symmetric linear case we set $\sigma=0$ and consider the linear Volterra-system associated with the initial condition given by
\be \label{ICB}
\psi^{L}_{beat,0}(x):=\alpha\,\phi_f(x)+\beta\, \phi_e(x)\,,  \quad \alpha,\beta\in \RR\,,
\ee
which can be exactly solved:

\[
\left\{
\begin{array}{l}
q_1(t)=\alpha\, \phi_f(-a)\, e^{\imath\,\lambda_f\,t} +\beta\, \phi_e(-a)\, e^{\imath\,\lambda_e\,t}\,, \\[3mm]
q_2(t)=\alpha\, \phi_f(a)\, e^{\imath\,\lambda_f\,t} +\beta\, \phi_e(a)\, e^{\imath\,\lambda_e\,t}\,, 
\end{array}
\right.
\quad \forall t \in \RR^+\,.
\]

\begin{figure}[htbp]
\begin{center}
\includegraphics[width=7cm]{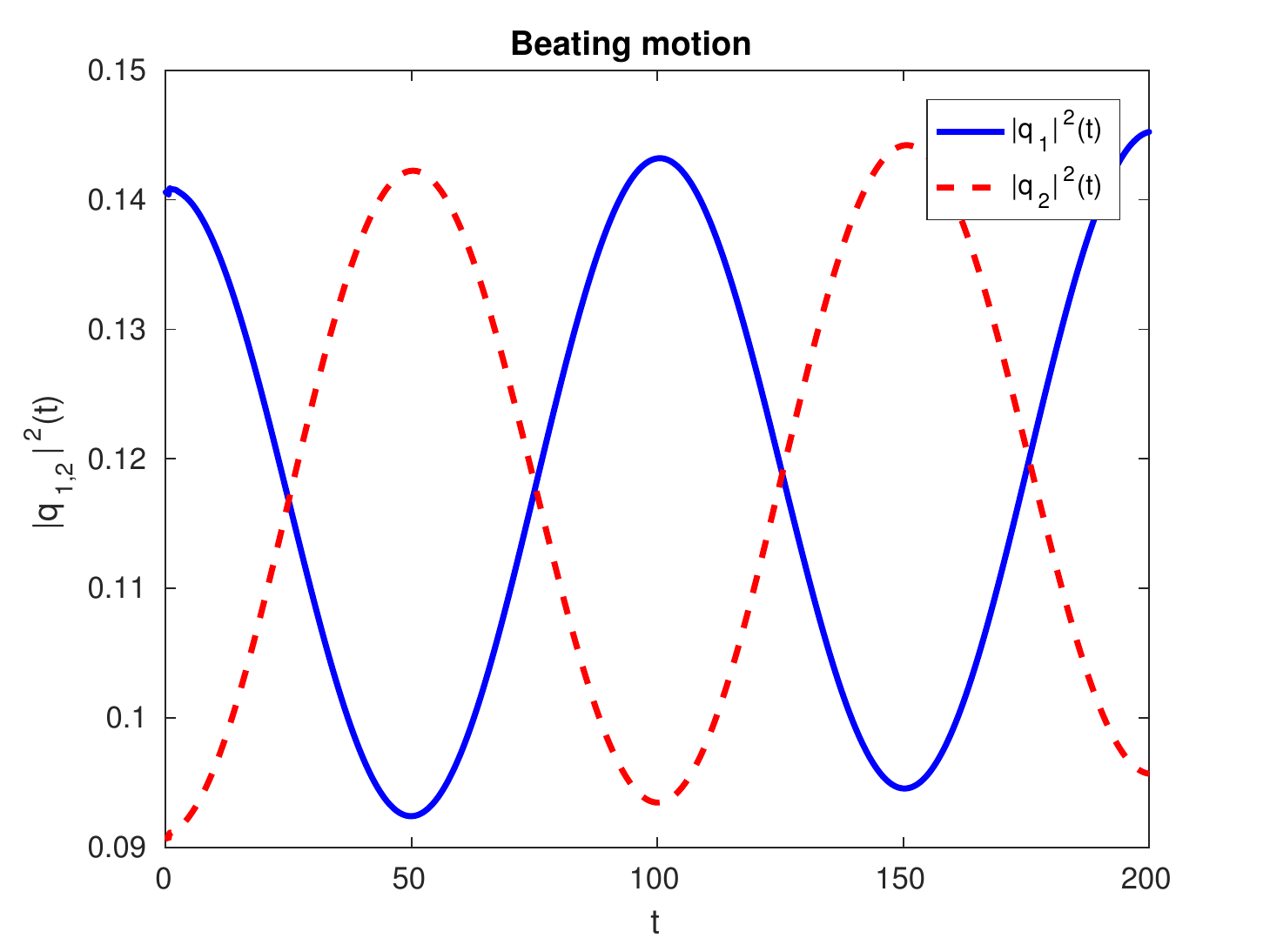}

\end{center}
\caption{\label{IMA0} 
{\footnotesize The beating effect.  Evolution in time of the numerical solutions $|q_1|^2(t)$ resp. $|q_2|^2(t)$  for $a=3$, $\alpha=\sqrt{0.01}$, $\beta=\sqrt{0.99}$, $\gamma=-0.5$.}}
\end{figure}

\noindent Figure \ref{IMA0} presents on the left the time-evolution of the numerical solutions of the Volterra-system  associated to the parameters indicated in the figure caption.

\begin{figure}[htbp]
\begin{center}
\includegraphics[width=5cm]{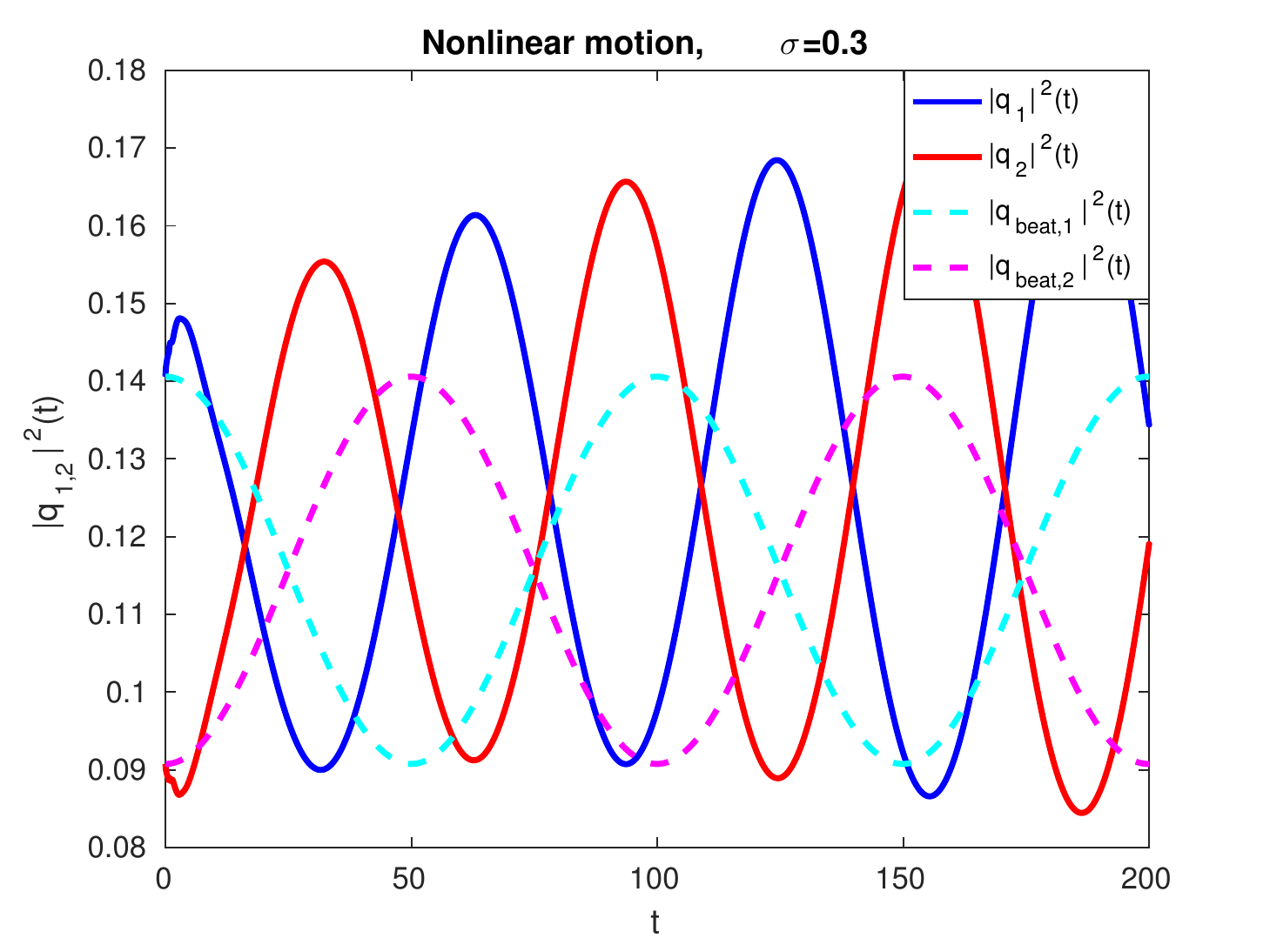}\hfill
\includegraphics[width=5cm]{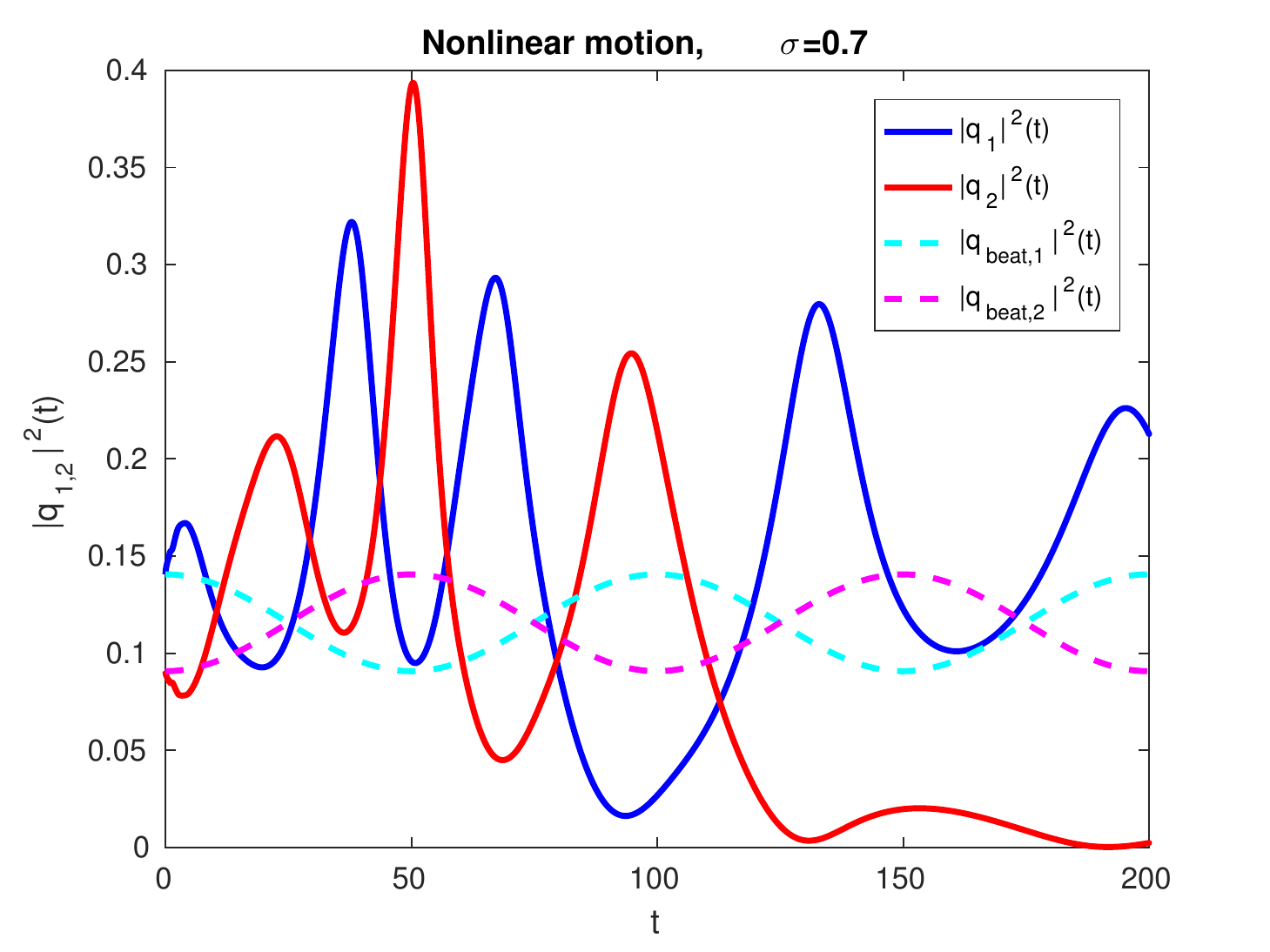}\hfill
\includegraphics[width=5cm]{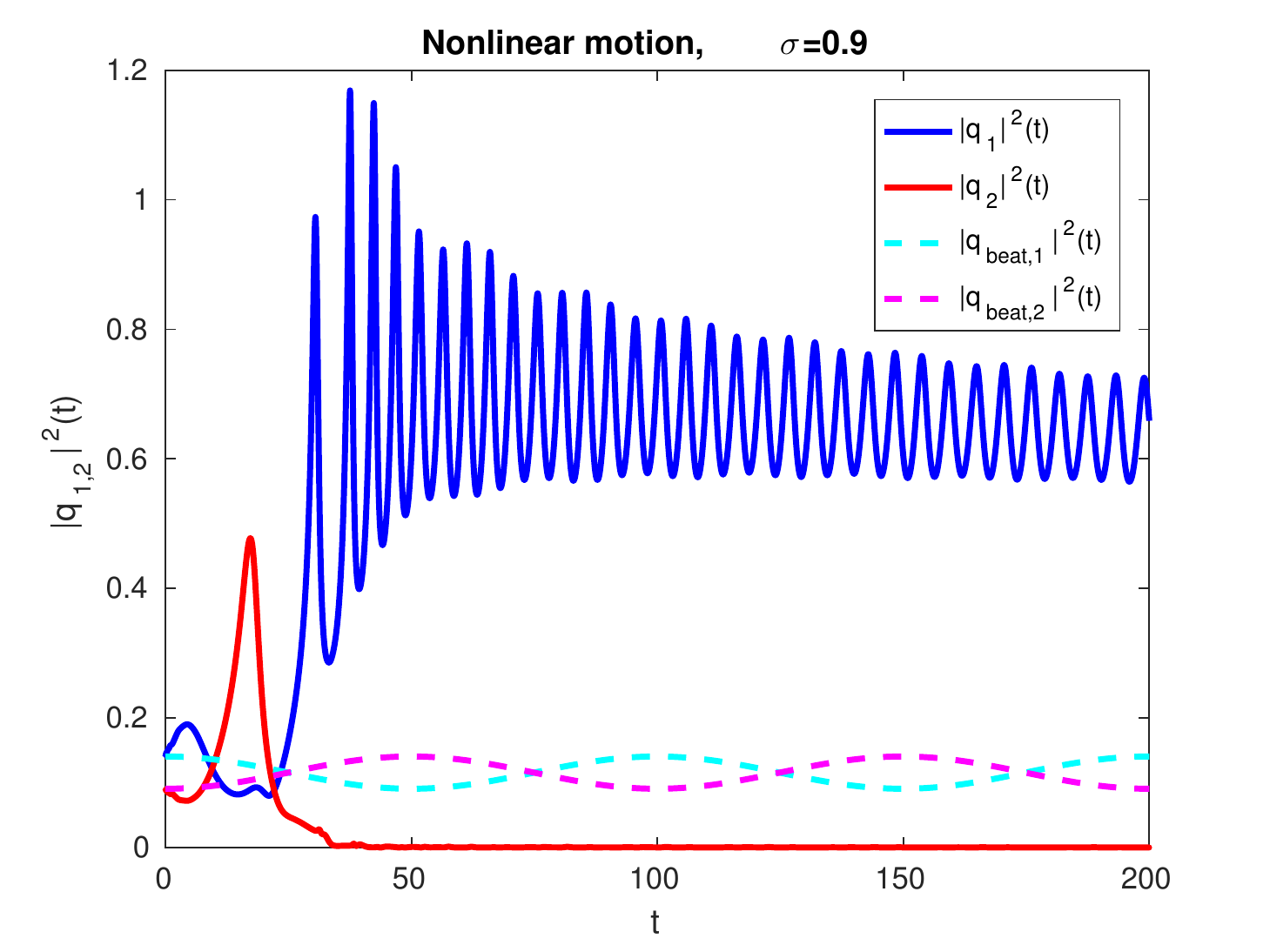}
\end{center}
\caption{\label{IMA1} 
{\footnotesize The non-linear time-evolution of the numerical solutions $|q_{1}^{num}|^2(t)$ resp. $|q_{2}^{num}|^2(t)$ (in blue/red full line) and corresponding linear beating solutions $|q_{beat,1}|^2(t)$ resp. $|q_{beat,2}|^2(t)$ (in cyan/magenta dashed line), for $\sigma=0.3$ (left), $\sigma=0.7$ (center)  and $\sigma=0.9$ (right).}}
\end{figure}

 Let us consider  the non-linear case. We assume  the same initial condition and the same parameters as in the symmetric linear case \eqref{ICB}. Increasing the power of the nonlinearity (we consider $\sigma=0.3,\,\sigma=0.7,\,\sigma=0.9$) we  observe that the time elapsed till the suppression of the beating effect is getting shorter and shorter.
Here the time dependent point interaction strength is 
$$
\gamma_\pm(t)=\gamma\, |\psi(t,\pm a)|^{2 \sigma}\,.
$$
In order to have at time $t=0$ the same strength of the linear case $\gamma_\pm(0)=-0.5$ we assume 
$$
\gamma:=2\,\gamma_\pm(0)/[|\psi_{0}(a)|^{2 \sigma}+|\psi_{0}(- a)|^{2 \sigma}]\,.
$$
In Figure \ref{IMA1} we plot the numerical solutions of the Volterra-system, {\it i.e.} $|q_{1}^{num}|^2(t)$ resp. $|q_{2}^{num}|^2(t)$ (in blue resp. red) as functions of time, and for the different non-linearity exponents. As a reference, we plot also the exact solutions of the symmetric linear system, {\it i.e.}  $|q_{beat,1}|^2(t)$ resp. $|q_{beat,2}|^2(t)$ (in cyan resp. magenta). Figures show clearly how the non-linearity suppresses the beating-effect.

\section{Conclusion}

In our numerical simulation we showed that, in a zero range non-linear double well potential,   the quantum beating mechanism is highly unstable under perturbations breaking the inversion symmetry of the problem.

The results shown in this review require further developments and extensions. In particular it is necessary to examine analytically the solutions of the system of Volterra integral equations (\ref{VOLT})
to clarify the dependence on initial conditions of the time needed for the beating suppression. A generalization of the results to higher dimensions is in progress.

\bigskip

\noindent {\bf Acknowledgments.} R.C. and L.T.  acknowledge the support of the FIR 2013 project ``Condensed Matter in Mathematical Physics'', Ministry of University and Research of Italian Republic  (code RBFR13WAET). C.N. would like to acknowledge support from the CNRS-PICS project ``MANUS'' (Modelling and Numerics of Spintronics and Graphenes, 2016-2018). 

\bibliographystyle{unsrt}


\end{document}